# Experimental Analysis of Safety Application Reliability in V2V Networks

Biplav Choudhury, Vijay K Shah, Avik Dayal, Jeffrey H. Reed

*Abstract*—Vehicle-to-Vehicle (V2V) communication networks enable safety applications via periodic broadcast of Basic Safety Messages (BSMs) or *safety beacons*. Beacons include time-critical information such as sender vehicle's location, speed and direction. The vehicle density may be very high in certain scenarios and such V2V networks suffer from channel congestion and undesirable level of packet collisions; which in turn may seriously jeopardize safety application reliability and cause collision risky situations. In this work, we perform experimental analysis of safety application reliability (in terms of *collision risks*), and conclude that there exists a unique beacon rate for which the safety performance is maximized, and the optimal beacon rate is unique for varying vehicle densities. The collision risk of a certain vehicle is computed using a simple kinematics-based model, and is based on *tracking error*, defined as the difference between vehicle's actual position and the perceived location of that vehicle by its neighbors (via most-recent beacons). Furthermore, we analyze the interconnection between the collision risk and two well-known network performance metrics, *Age of Information* (AoI) and *throughput*. Our experimentation shows that AoI has a strong correlation with the collision risk and AoI-optimal beacon rate is similar to the safety-optimal beacon rate, irrespective of the vehicle densities, queuing sizes and disciplines. Whereas throughput works well only under higher vehicle densities.

## I. INTRODUCTION

Communication in Vehicle-to-Everything (V2X) and Vehicle-to-Vehicle (V2V) networks has been a topic of active research due to its potential to reduce vehicle collisions and other applications. The two standards established for it are Dedicated Short Range Communications (DSRC) and Cellular V2X (C-V2X). DSRC is based on 802.11 WiFi-based technology that establishes a decentralized network. DSRC relies on the 802.11p standard, though efforts are ongoing to come up with a better version called 802.11bd [1]. The other standard is Cellular-V2X (C-V2X) which was developed by 3$^{\text{rd}}$ Generation Partnership Project (3GPP) in its Rel-14. It is a cellular based technology with additional transmission modes that uses the sidelink interface PC5 for direct V2X communications, and the newer version is called New-Radio V2X (NR-V2X) [1]. Currently, the three major V2X use cases are: basic safety messages, cooperative traffic efficiency messages and media sharing applications.

- Basic Safety Messages (BSMs) or *safety beacons*: Periodically broadcast low latency messages ($\leq 100$ ms) that carry information about vehicle position, speed etc.
- Cooperative traffic efficiency messages: Event triggered low latency ($\leq 200$ ms) messages initiated by traffic management systems designed to improve traffic flow.
- Media Sharing Applications: Peer-to-peer Infotainment content is shared among nearby vehicles.

Basic Safety Messages (BSMs) or *safety beacons* enable several safety applications like Forward Collision Warning, Blind Spot/Lane Change Warning, Intersection Movement Assist etc. and enhances road safety by greatly minimizing the collision risky situations[1]. This is because safety beacons facilitate accurate positioning or *localization* of neighboring vehicles. In such V2V networks, it is imperative that the safety beacons are not lost due to packet collisions due to congestion, and that the beacons are up-to-date and *fresh*. Lost or outdated safety beacons will negatively impact the reliability of safety applications as it will lead to increased occurrence of *collision risky* situations. Collision risk increases due to wrong localization of neighboring vehicles or in other words, high *tracking error*. Tracking error is the difference between a vehicle's actual location and perceived location of that vehicle by its neighbors (via safety beacons).

*Safety beacon rate*, i.e., the rate at which beacons are broadcast, becomes a critical parameter in order to ensure minimal channel congestion and improved reliability of safety applications. In 802.11p DSRC standard, the beacon rate is set to 10 Hz, which may not work well in all V2V network scenarios, as also pointed out by existing works [2]. Consider the following two instances: (i) high vehicle density scenarios - the V2V network would likely suffer from high channel congestion and undesirable level of packet collisions even when beacon rate is as low as 10 Hz, and (ii) low vehicle density scenarios - the reliability of safety applications can be further enhanced with increased beacon rate (above 10 Hz), as higher beacon rate would improve the freshness of the information and packet collisions do not worsen (as the channels can support higher beacon rate for fewer vehicles). Refer to Section II for a detailed overview of related works.

In this paper, we perform a comprehensive investigation of the safety applications reliability against varying safety beacon rates and vehicle densities through experimental analysis based on ns-3 network simulator[2]. We utilize *Collision risk* as the metric for safety application reliability, which is computed using a simple kinematics-based model based on tracking error. We observe that there exists an optimal beacon rate for which the collision risk is minimized, and the optimal value is unique for varying vehicle densities. For instance, in our experiments, the safety-optimal beacon rate for 50 vehicles/km is 25 Hz whereas it is 10 Hz for 200 vehicles/km.

Furthermore, we investigate two well-known network metrics, namely *Age of Information* (AoI) [3] and *Throughput*, and

This research is supported by the Office of Naval Research (ONR) under MURI Grant N00014-19-1-2621. The authors are with Bradley Department of Electrical and Computer Engineering, Virginia Tech, and affiliated with Wireless@VT Lab. (Email:{biplavc, vijays, ad6db, reedjh}@vt.edu)

---

[1] Additional sensors like camera, radars are also being increasingly used for safety application purpose

[2] https://www.nsnam.org/

analyze their relation with safety performance, i.e., collision risk. AoI is a recently conceived network metric that captures the freshness of information, which is of greater interest to time-sensitive V2V networks. AoI is defined as the time elapsed since the last beacon was received at a certain vehicle from the beacon's generation time. On the contrary the second metric, *throughput*, is defined as the rate of data transfer and has been studied extensively in various wireless networks [4]. Our results shows that AoI-optimal beacon rate has a strong correlation with the safety-optimal beacon rate, irrespective of varying vehicle densities and queuing considerations. Whereas throughput-optimal rate works well for safety only at higher vehicle densities. The results also indicate that longer queues lead to high queuing delay and packet loss, and thus a queue size of 1 is both AoI and throughput optimal in all scenarios.

The paper is organized as follows. Sec. II discusses the related works. Sec. III describes the tracking error and collision risk whereas Sec. IV presents the two network metrics, i.e., AoI and throughput. Sec. V details the experimental analysis. Finally, Sec.VI concludes the paper.

## II. LITERATURE REVIEW

This section discusses the related works on safety performance, AoI and throughput in the context of V2V networks.

In [5], the authors analyzed the effect of beacon rate on network performance of safety messages in V2V networks, and developed a framework to recommend beacon rates based on the message utility maximization. Remote mobility estimators were used in [6] and [7] to control tracking error using adaptive beacon rates. The authors used Channel Busy Ratio (CBR) for controlling beacon rate in [8] to improve safety information dissemination range. [9] and [10] define risk metrics based on the vehicle dynamics and traffic situation, which are then used to select beacon rates. In [11], an adaptive beacon rate algorithm was designed to improve safety in terms of beacon inter-reception time. All these works focus on improving safety performance via controlling beacon rates. On contrary, this paper investigates the relationship of safety performance with varying vehicle densities and queue sizes and disciplines, in addition to beacon rates. Furthermore, we investigate the interconnection of safety application reliability with two network metrics, i.e., AoI and Throughput.

AoI was introduced in [3] and a distributed algorithm for rate adaptation to reduce AoI was proposed in it. In [12], the authors develop an analytical model that takes into account the Carrier Sense Multiple Access (CSMA) contention to improve the AoI in a V2V network. Llatser et al. [13] simulated a convoy of automated vehicles where AoI was analyzed along with other parameters for changing convoy-size and beacon rates. A centralized scheduling scheme for beacon broadcasting was designed in [14] to minimize AoI and it was shown to outperform DSRC in terms of fairness and number of effective neighbors. Other papers that deal with AoI can be found in [15]. On the other hand, throughput is a classical metric and it has been well studied for various networks. [4] lists several of the works that analyses throughput in V2V networks. However, none of these works investigate their interconnection with safety performance, which is a key requirement to minimize collision risky situations.

## III. TRACKING ERROR AND COLLISION RISK

The calculation of the tracking error and the resulting collision risk is explained below:

### A. Tracking Error

Consider a basic transmitter vehicle and a receiver vehicle that sends and receives beacons, respectively. Let the most recent beacon from sender $u$ to receiver $v$ carry the information that $u$'s location is $x_{uv}^{t_1'}$, which is true at the packet generation time $t_1'$. At time $t_1$ when the beacon is received, the position of $u$ has changed to $x_u^{t_1}$. The tracking error at $v$ while tracking $u$ at time $t_1$ with a reception is [7]:

$$\delta pos_{uv}^{t_1} = |x_u^{t_1} - x_{uv}^{t_1'}|, \quad \text{where} \quad t_1' < t_1 \quad (1)$$

We consider x-coordinate to calculate the tracking error as the vehicles move in x-direction only. See Sec.V-A for details. The error in Eqn.1 occurs because (i) the beacon takes non-zero time to be delivered to the receiver after being generated, and (ii) the sender would likely move a certain distance during this time period. For large time delays, the beacon's information is relatively outdated. Packet losses will further deteriorate this error. If there were beacon receptions at instants $t_1$ and $t_2$ whose generation times are $t_1'$ and $t_2'$ respectively, the errors at times $t_1$ and $t_2$ are calculated using Eqn.1. However, at time instant $t_1 < \widetilde{t} < t_2$, where were no beacon receptions, we compute the tracking error at vehicle $v$ at time $\widetilde{t}$, as follows:

$$\delta pos_{uv}^{\widetilde{t}} = |x_u^{\widetilde{t}} - x_{uv}^{t_1'}| \quad (2)$$

where $x_u^{\widetilde{t}}$ is the location of vehicle $u$ at time instant $\widetilde{t}$, which can be computed as the product of vehicle u's speed and time difference $|\widetilde{t} - t_1|$. $x_{uv}^{t_1'}$ is the location of $u$ as per the last beacon received at vehicle $v$ from vehicle $u$ at $t_1$. Therefore the tracking error is calculated using Eqn.1 at instants of beacon reception and Eqn.2 at instants of no reception.

In the observation interval $T$, the average tracking error for a certain vehicle pair $u - v$, is as follows:

$$\delta pos_{uv} = \frac{1}{T} \int_T \delta pos_{uv}^t \quad (3)$$

where $t$ represents any instant in the observation interval $T$ including both the instants of reception and non-reception.

### B. Collision Risk

Time To Collision (TTC) for a pair of vehicles is defined as the time taken for the distance between the pair of vehicles to become 0, which denotes a possible collision between them. We relate the collision risk to the error in the TTC, which results from the tracking error. The collision scenario is shown in Fig. 1. Every vehicle continuously monitors the TTC with respect to its neighbors as it is a key indicator of safety application reliability [9]. High TTCs mean that the neighboring vehicles are far apart and there is no immediate threat of collision and vice versa.

At any time $t$, the receiver $v$ calculates TTC with respect to its neighbor $u$ based on the beacon sent by $u$ as:

$$TTC_{uv,calc}^t = |\frac{1}{s_{uv}}(x_v^t - x_{uv}^{t'})| \quad (4)$$

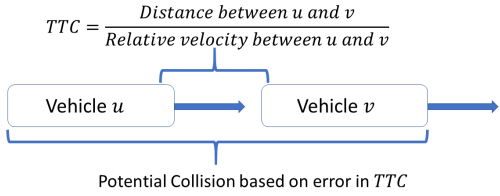

Fig. 1: TTC Calculation

$x_v^t$ is $v$'s actual position at time $t$ and $x_{uv}^{t'}$ is $u$'s location as per the last beacon received at $v$ from $u$ with generation time $t'$. $s_{uv}$ is the relative velocity between them.

However, the actual TTC at time $t$ is:

$$TTC_{uv,act}^t = |\frac{1}{s_{uv}}(x_v^t - x_u^t)| \quad (5)$$

as at $t$, sender $u$ is at $x_u^t$ and not $x_{uv}^{t'}$. So the tracking error affects the TTC calculation. This TTC error at any time $t$ can be calculated as the absolute value of the difference between Eqn.4 and Eqn.5, and is equal to $\delta TTC_{uv}^t = \frac{\delta pos_{uv}^t}{s_{uv}}$. In the observation interval $T$, the average TTC error for $u$-$v$ is:

$$\delta TTC_{uv} = \frac{1}{T}\int_T \delta TTC_{uv}^t = \frac{\delta pos_{uv}}{s_{uv}} \quad (6)$$

This $\delta TTC_{uv}^t$ can lead to risky situations and its significance in collision warning was studied in [16]. E.g. suppose $v$ has overestimated/underestimated[3] the TTC to $u$ and is about to take a route/maneuver decision based on this erroneous value of TTC. Such a situation can be safely averted by manual intervention if the time taken by the driver to react and apply the brakes to make the vehicle stop is less than $\delta TTC_{uv}^t$. But if this error exceeds the driver's controllability, it presents a risky scenario. We classify the scenario to be risky if:

$$\delta TTC_{uv} > t_{react} + t_{brake} \quad (7)$$

We use a Gaussian distribution to model the speeds of the vehicles with a mean of 25 m/s and a variance of 3 m/s [17]. $t_{react}$ is taken to be 1s [9] and is the time taken by the driver to respond to the situation and apply the brakes. We take the common deceleration of all the vehicles as 4.6 m/s$^2$ [9], and this leads to a value of $t_{brake} = 5.43$ s for the vehicles to come to a complete stop. Eqn. 7 then becomes: $\delta TTC_{uv} > 6.43$ s for a scenario to be risky.

For every simulation performed, we count the proportion of the vehicle pairs whose $\delta TTC_{uv}$ exceeded 6.43 s during the interval they communicated. This is taken as the measure of the overall safety performance of the network. The logic is that if $\delta TTC_{uv}$ was under threshold for a particular pair, the receiver's routing/maneuver decisions would have considered the correct location of the sender most of the times. So it is unlikely that the receiver's actions will cause any collisions. It should be kept in mind that the TTC error arises solely due to the tracking error, which in turn arises due to the delay and packet loss in the communication. Thus the safety performance is being measured from a purely communications point of view, and no other phenomenon is considered.

[3]We consider both overestimating and underestimating TTC to be hazardous as overestimating means the vehicles get too close, and decisions based on underestimated values will impact the other neighbors

## IV. AGE OF INFORMATION AND THROUGHPUT

In this section, we briefly discuss two network metrics, i.e., Age of Information (AoI), and throughput. In the following Section V, we study their interconnection with safety performance, i.e., collision risk, under various V2V scenarios.

### A. Age of Information

Age of Information (AoI) at the receiving vehicles is calculated as the time elapsed since the received beacon was generated at the sender until the beacon's reception time. It is equal to the end-to-end delays at the instants of reception, with linear increase between receptions. While delay is associated with the packets and is undefined when there is no reception, AoI can be thought of as a continuous variable that is defined for the entire observation interval. We define the observation interval $T$ as the interval between the reception of the first and the last beacons between a sender and a receiver. The average AoI is calculated as the total area under the age plot normalized by the observation interval [3], i.e. $\triangle_{uv} = \frac{1}{T}\int_T \triangle_{uv}^t$, where $\triangle_{uv}^t$ represents the AoI of vehicle $u$'s information at vehicle $v$. It is calculated as $\triangle_{uv}^t = t - g(t)$ at instant $t$, with $g(t)$ being the generation time of the most recent packet received by $v$. The system AoI with $N$ vehicles is calculated across $N(N-1)$ unique pairs of sender and receiver as:

$$\triangle = \frac{1}{N(N-1)}\sum_u \sum_{v \neq u} \triangle_{uv} \quad (8)$$

As our work is focused on safety applications, we don't consider Last Come First Serve (LCFS) queues of sizes more than 1, as having a bigger LCFS queue provides no practical utility in terms of time-sensitive information. This is because an LCFS queue will send out the most recently generated packet first making the older packets wait, and these older packets do not add any value to the receiver vehicle after the newer packet has been received. They are differentiated as *informative* and *non-informative* packets [18]. However for the sake of completeness, we show the variation of AoI in a specific case with LCFS queues in Fig. 2. It can be seen that in LCFS, queue sizes do not make a big difference in AoI. In the rest of our analysis, we won't be showing the LCFS results as LCFS of the considered queue sizes (1,5,10,100 packets) perform very close to a First Come First Come (FCFS) queue of size 1 as per Fig.2[4]. Therefore, the AoI performance of an LCFS queue can be roughly assumed to have a performance similar to that of an FCFS for queue size 1. The reception of non-informative packets do not lead to any reduction in AoI [18]. This can also be understood in terms of tracking error: once the recent location of a vehicle is known via a recent beacon, the location information obtained from the older beacons that are transmitted after the newer beacon in an LCFS queue do not add any value. Tracking error for any LCFS queue then becomes similar to FCFS queue's tracking error of queue size 1. As collision risk is computed directly from the tracking error, the above property of LCFS queue applies to collision risk in LCFS queues too.

[4]a small but negligible AoI difference between queue sizes of 1 and 100 can be noticed for 200 veh/km.

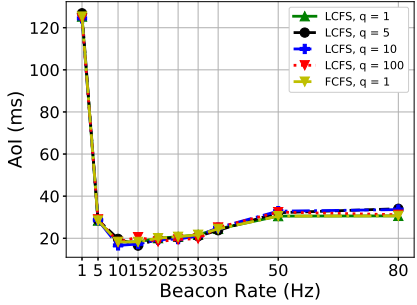

Fig. 2: Age in an LCFS queue

## B. Throughput

Throughput is a measure of the rate of data transfer, and it is calculated as the total data transferred in the entire simulation. Taking $\tau$ as the observation interval and $K$ as total packets transferred in that duration, total throughput = $\frac{K}{\tau}$ packets per second. Average throughput per vehicle can be obtained by normalizing the total throughput with the number of vehicles. Throughput is not affected by LCFS/FCFS, and the performance of an LCFS queue would be exactly same as than of a FCFS queue of the same size.

## V. EXPERIMENTAL ANALYSIS

This section describes the simulation setup, followed by performance analysis of safety application reliability, i.e., collision risk and two network metrics, i.e., AoI and Throughput, against varying beacon rates and vehicle densities.

### A. Simulation Setup

We utilize Network Simulator 3 (ns-3) for our experimental study. We consider that the vehicles are placed uniformly[5] along a 3-lane highway of length 1 Km. The vehicles move along the highway at constant speeds along the x-axis without any lateral movement across the lanes, until the simulation ends. Unless otherwise stated, the beacon/transmission rates were varied from 1 Hz to 80 Hz, and a small random number is added to this rate to ensure that all the vehicles do not attempt transmission at the same time [3]. The packet sizes are set to 320 bytes The queue sizes are varied between 1, 5, 10 and 100 packets, and the queue discipline is the FCFS. Recall LCFS is equivalent to FCFS with size 1, irrespective of queue sizes for LCFS. A list of the important parameters used in the simulation along with their values are given in Table I.

TABLE I: Parameters used to set up the simulation

| Parameters | Value |
| --- | --- |
| Vehicle Density | 50 veh/km - 200 veh/km |
| Number of Lanes | 3 |
| Lane Width | 4 m |
| Packet Size | 320 bytes |
| Data Rate | 6 Mbps |
| Loss Model | Log Distance Propagation Loss |
| Path Loss exponent | $\gamma = 3$ |
| Channel Frequency | 5.9 GHz |
| Channel Bandwidth | 10 MHz |
| Vehicle Speeds | $N(25 \text{ m/s}, 3 \text{ m/s})$ |
| $t_{brake}$ | 5.43 s |
| $t_{react}$ | 1 s |
| Antennas/Spatial Streams | 1/1 |

### B. Experimental Results

In the following, we discuss the comparative analysis of Collision risk, AoI, and Throughput against varying beacon rates and vehicle densities.

---

[5]This uniform placement of vehicles do not affect the results as the average measures for random and uniform placement are very similar [8]

**Collision Risk Analysis:** Fig. 3 depicts how collision risk varies with varying beacon rates, under considered vehicle densities of 50 vehicles/Km and 200 vehicles/Km. Collision risk is minimal at 25 Hz for 50 veh/km whereas 10 Hz for 200 veh/km, which means that there is a unique beacon rate for a certain vehicle density that maximizes the safety application reliability. Furthermore, note that there exists a convex relationship between the collision risk and beacon rates, which hints that gradient-descent algorithms may be utilized to analytically determine the optimal safety-optimal beacon rate. (We will investigate this as a part of our future work.) From this analysis, it is clear that enforcing a fixed beacon rate of 10 Hz (by DSRC standard) may not be optimal in all V2V scenarios. Regarding the queuing sizes, collision risk remains unaffected for lower beacon rates, however, it increases gradually after 35 Hz for lower vehicle density, and after 25 Hz for higher vehicle density. Longer queues perform poorly compared to that of shorter queues, and interestingly, queue size 1 performs the best. These observations can mainly be credited to its correlation with AoI (next paragraph).

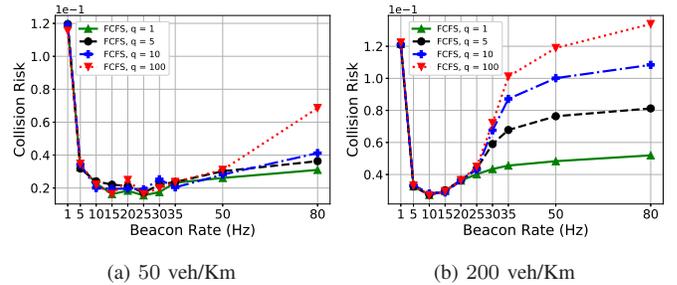

(a) 50 veh/Km  (b) 200 veh/Km

Fig. 3: Collision risk vs Beacon rates

**AoI Analysis:** Fig.4 depicts the relationship between AoI and safety beacon rates under varying vehicle densities. Similar to collision risk, it is interesting to see that AoI is also a convex function of beacon rate. More interestingly, the AoI-optimal beacon rate is the same as the safety-optimal beacon rate, for both the considered vehicle densities. This is an important result as it establishes that AoI and safety application reliability are highly correlated, and safety application's beacon rate algorithm can reliably undertake AoI minimization as the sole objective, rather than safety application reliability[6].

Similar to collision risk, AoI also remains unaffected for lower beacon rates, and after optimal beacon rate, it increases gradually for both the density scenarios. Smaller queues have lower AoI and this can be largely attributed to lower queuing delays. Fig. 5 shows that delays are lower for shorter queues, which in turn results in lower AoI and vice-versa. At very high beacon rates, the network becomes saturated, which results in a saturation in AoI, as also pointed out in the literature [12]. An important observation is that at lower beacon rates, delay is less (Fig.5) as the channel resources are available whereas the collision risk is very high (Fig. 3). This shows that lower delay doesn't guarantee higher safety performance. However similar to collision risk, AoI is higher for both lower and higher beacon rates. AoI is higher for lower beacon rates due to the high inter-arrival times while at higher beacon rates, AoI is higher mainly due to higher queuing delays.

---

[6]Computing safety application reliability, i.e., collision risk, in real-time is very hard in V2V networks compared to that of AoI, a network metric.

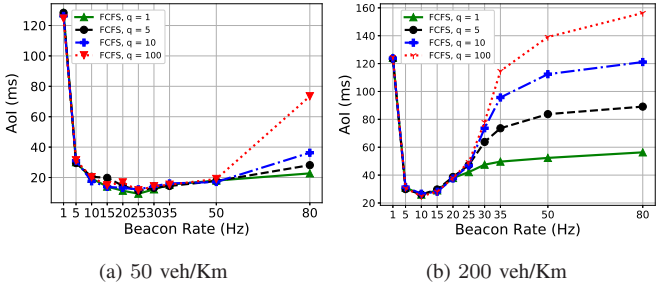

(a) 50 veh/Km  (b) 200 veh/Km

Fig. 4: Age of Information vs Beacon rates

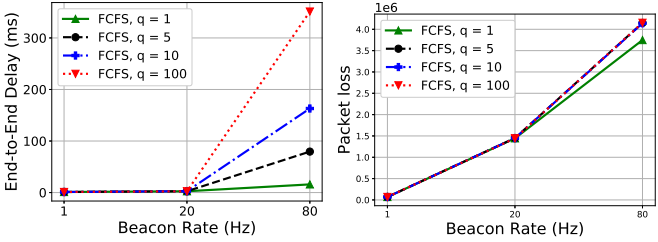

Fig. 5: End-to-end-delay    Fig. 6: Packets lost in collisions

**Throughput Analysis:** Fig. 7 shows the gradual increase in throughput until it peaks, and then it saturates to a lower value. For 200 veh/Km, it peaks at 15 Hz and starts saturating from 35 Hz; while it doesn't saturate until 80 Hz for 50 veh/Km.

Notice that for lower vehicle density, the throughput-optimal beacon rate (80 Hz) is very far from safety-optimal rate (25 Hz), however, for higher vehicle density, throughput-optimal beacon rate (15 Hz) is closer to the safety-optimal rate (10 Hz), and performs quite well in terms of collision risk. Similar to collision risk and AoI, a queue size of 1 is throughput-optimal, and this can be attributed to the fact that packet loss due to packet collisions is less for smaller queues, compared to that of larger queues [19] (See Fig. 6).

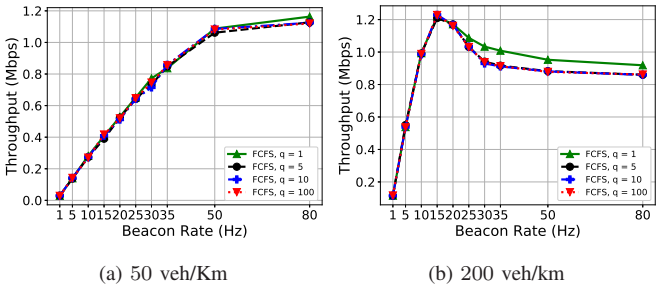

(a) 50 veh/Km  (b) 200 veh/km

Fig. 7: Throughput vs Beacon rates

## VI. CONCLUSIONS

In this paper, we performed a comprehensive experimental analysis of safety application reliability (in terms of collision risks) against beacon rates and vehicle densities for V2V networks, and concluded that there exists a unique beacon rate for which the safety application reliability is maximized, and it is unique for each vehicle density. Furthermore, our experiments showed that there exists a strong correlation between collision risk and Age of Information (AoI) notwithstanding vehicle densities, and other queuing parameters. Whereas, the throughput-optimal rate approaches safety-optimal beacon rate only under higher vehicle densities. Queue size of 1 is both AoI and throughput optimal in all considered V2V scenarios.

In the future, we would like to investigate whether there exists a mathematical relationship between AoI and collision risk, so that an optimization technique, like, gradient descent based technique, can be utilized to compute optimal beacon rates for undertaken V2V scenario. Additionally, we would propose a novel adaptive beacon rate algorithm that minimizes AoI (and thus, enhances safety application reliability), while considering vehicle densities and queuing considerations.